\documentclass[12pt,cite,epsf,epsfig]{article}
\usepackage[usenames,dvipsnames]{pstricks}
\usepackage{epsfig}
\usepackage{epstopdf}
\usepackage{amsmath}
\usepackage{amssymb}
\usepackage{amsfonts}

\begin{document}
\title{Prospects for Reconstruction of Leptonic Unitarity Quadrangle and Neutrino Oscillation Experiments}
\author {Surender Verma\thanks{Electronic address: s\_7verma@yahoo.co.in}  and Shankita Bhardwaj\thanks{Electronic address: shankita.bhardwaj982@gmail.com}}

\date{\textit{Department of Physics and Astronomical Science,\\Central University of Himachal Pradesh, Dharamshala 176215, INDIA.}}
\maketitle

\begin{abstract}
After the observation of non-zero $\theta_{13}$ the goal has shifted to observe $CP$ violation in the leptonic sector.
Neutrino oscillation experiments can, directly, probe  the Dirac $CP$ phases. Alternatively,
one can measure $CP$ violation in the leptonic sector using Leptonic Unitarity Quadrangle(LUQ).
The existence of Standard Model (SM) gauge singlets - sterile neutrinos - will provide additional
sources of $CP$ violation. We investigate the connection between neutrino survival probability and
rephasing invariants of the $4\times4$ neutrino mixing matrix. In general, LUQ contain eight geometrical
parameters out of which five are independent. We obtain $CP$ asymmetry($P_{\nu_f\rightarrow\nu_{f'}}-P_{\bar{\nu}_f\rightarrow\bar{\nu}_{f'}}$)
in terms of these independent parameters of the LUQ and search for the possibilities of extracting
information on these independent geometrical  parameters in short baseline(SBL) and long baseline(LBL)
experiments, thus, looking for constructing LUQ and possible measurement of $CP$ violation.
We find that it is not possible to construct LUQ using data from LBL experiments because $CP$
asymmetry is sensitive to only three of the five independent parameters of LUQ. However, for SBL
experiments, $CP$ asymmetry is found to be sensitive to all five independent parameters making it
possible to construct LUQ and measure $CP$ violation.
\end{abstract}

\section{Introduction}
The observation of non-zero $\theta_{13}$ \cite{dc, db, reno} has, conclusively, established three types of 
oscillations and provides an opportunity for possible measurement of i) $CP$ violation in the leptonic
sector, and ii) to determine neutrino mass hierarchy. $CP$ violation is, essentially, a three flavor effect
\cite{cp} and is attributed to the non-trivial phases of the neutrino mixing matrix. In general, for three 
generation case, the neutrino mixing matrix contain three phases, one Dirac-type $CP$ violating phase and
two Majorana-type $CP$ violating phases. However, only Dirac phase manifests itself in the neutrino
oscillation experiments. Looking at the current and future neutrino facilities, which are either at
planning or data acquisition stage, the measurement of $CP$ violation is not beyond realization
\cite{t2k, minos}. The current experimental data on neutrino masses and mixings can be explained
within the paradigm of three active neutrino isodoublets. However, reconciliation with the short
baseline(SBL) anomalies such as LSND, MiniBooNE and Gallium \cite{lsnd, mb, mb1, gm, tm, df} requires
the introduction of Standard Model(SM) gauge singlet(s)-sterile neutrino(s) because they involve the
mass-squared difference, $\Delta m^{2}_{\text{SBL}}\gg\Delta m_{\text{Atm}}^{2}\gg\Delta m_{\text{Solar}}^{2}$.
The possible existence of Standard Model(SM) gauge singlet fermion(s) is an attractive extension to our quest
to understand fundamental physics including origin of non-zero neutrino masses and dark matter puzzle.
In presence of sterile neutrinos, the standard three neutrino picture must be enlarged to accommodate
the more mass eigenstates having non-zero mixing with the standard three active flavors. Also,
there will be additional sources of $CP$ violation in presence of these SM gauge singlets. So,
it is important to study the prospects of detecting these additional sources of $CP$ violation.
In general, $CP$ violation can be studied in two ways. One way is to directly measure $CP$ violating
phase in the neutrino oscillation experiments \cite{t2k,minos} and second is to construct the leptonic
unitarity triangle(LUT)/quadrangle(LUQ) \cite{he, xing, xing1}. In the present work, we have followed the second approach.
In four neutrino mixing models, $CP$ violation is, generally, expected to be violated and is attributed
to the nontrivial complex phases in $4\times 4$ neutrino mixing matrix. Neutrino oscillation experiments
play crucial role to study Pontecorvo-Maki-Nakagawa-Sakata(PMNS) matrix and to, directly, measure $CP$
violation, $P_{\nu_f\rightarrow\nu_{f'}}-P_{\bar{\nu}_f\rightarrow\bar{\nu}_{f'}}\neq 0$, in the leptonic
sector. Alternatively, in order to measure $CP$ violation, in a rephasing invariant manner using Leptonic
Unitarity Quadrangle(LUQ), one has to construct rephase invariants from $4\times4$ neutrino mixing matrix
$V$ given by $J_{ff'}^{ij}\equiv \Im\left(V_{fi}V_{f'j}V_{fj}^{*}V_{f'i}^{*}\right)$ \cite{cj}, where $(i,j)=0,1,2,3$
and $(f,f')=s,e,\mu,\tau$. In Sec. II, we present the connection between neutrino survival probability
and rephasing invariants of the $4\times4$ neutrino mixing matrix. In general, LUQ contain eight geometric
parameters, out of which five parameters are independent. In Sec. III, we present $CP$ asymmetry,
$P_{\nu_f\rightarrow\nu_{f'}}-P_{\bar{\nu}_f\rightarrow\bar{\nu}_{f'}}$, in terms of these independent parameters
of the LUQ and search for the possibilities of extracting information on these independent geometrical  parameters
in short baseline(SBL) and long baseline(LBL) experiments, thus, looking for constructing LUQ and possible measurement
of $CP$ violation. In Sec. IV, we draw our conclusions.
\section{Connecting Leptonic Unitarity Quadrangle to Mixing matrix}
In four neutrino mixing, the flavor($\nu_f$, $f=s,e,\mu,\tau$) and mass eigenstates($\nu_j$, $j=0,1,2,3$) are connected
through (4$\times$4) unitary matrix $V$ as
\begin{equation}
\centering
\begin{pmatrix}
  \nu_{s}\\
  \nu_{e}\\
  \nu_{\mu}\\
  \nu_{\tau}
\end{pmatrix}
=
 \begin{pmatrix}
  V_{s0} & V_{s1} &V_{s2} & V_{s3}\\
  V_{e0} & V_{e1} &V_{e2} & V_{e3}\\
  V_{\mu0} & V_{\mu1} & V_{\mu2} & V_{\mu3}\\
  V_{\tau0} & V_{\tau1} & V_{\tau2} &V_{\tau3}\\
\end{pmatrix}  
\begin{pmatrix}
  \nu_{0}\\
  \nu_{1}\\
  \nu_{2}\\
  \nu_{3}
\end{pmatrix}.
\end{equation}
The unitarity of mixing matrix $V$ ($V^{\dagger}V=VV^{\dagger}=1$) provide eight normalization 
and twelve orthogonality relations which corresponds to twelve unitarity quadrangles in the complex plane.
Now, let's consider a flavor state $|\nu_f\rangle$ that converts to $|\nu_{f'}\rangle$ after travelling a 
distance $L$ km. The vacuum transition probability for this conversion is given by \cite{xing, smb}
\begin{equation}
 P_{\nu_f\rightarrow \nu_{f'}}=\delta_{ff'}-4\sum_{i<j}^{}\left(\Re\left(V_{fi}V_{f'j}V_{fj}^{*}V_{f'i}^{*}\right)\sin^2\left(X_{ij}\right)\right)
 +2\sum_{i<j}^{}\left(J_{ff'}^{ij}\sin\left(2X_{ij}\right)\right),
\end{equation}

where $ X_{ij}\equiv 1.27\Delta m_{ij}^{2}L/E$ with $\Delta m_{ij}^{2}\equiv m_j^{2}-m_i^{2}$,
$L$ is baseline length, and $E$ is the neutrino beam energy. The vacuum transition 
probability for antineutrino can be directly obtained from Eqn.(2) by replacing $J_{ff'}^{ij}\rightarrow-J_{ff'}^{ij}$. 
 The flavor transition can be attributed to change in phase shift \cite{smb, jb} of the transition probability
 by phase angle $\lambda_{f'f:ij}$ defined as 
\begin{equation}
\lambda_{f'f:ij}=arg\left(V_{f'i}V_{fi}^{*}V_{fj}V_{f'j}^{*}\right),
\end{equation}
with $\lambda_{f'f;ij}=-\lambda_{ff'ij}=-\lambda_{f'f;ji}$ and $X_{ij}=-X_{ji}$. In terms of phase
angle Eqn.(2) can be written as 
\begin{eqnarray}
\nonumber
 P_{\nu_f\rightarrow \nu_{f'}}= &\delta_{ff'}&-4\sum_{i<j}^{}\left(\Re\left(V_{fi}V_{f'j}V_{fj}^{*}V_{f'i}^{*}\right)\sin^2\left(X_{ij}-\lambda_{f'f;ij}\right)\right)\\
                           &+&2\sum_{i<j}^{}\left(J_{ff'}^{ij}\sin(2X_{ij}-\lambda_{f'f;ij})\right)
\end{eqnarray}
for neutrinos and 
\begin{eqnarray}
\nonumber
 P_{\bar\nu_f\rightarrow \bar\nu_{f'}}=&\delta_{ff'}&-4\sum_{i<j}^{}\left(\Re\left(V_{fi}V_{f'j}V_{fj}^{*}V_{f'i}^{*}\right)\sin^2\left(X_{ij}+\lambda_{f'f;ij}\right)\right)\\
                                   &-&2\sum_{i<j}^{}\left(J_{ff'}^{ij}\sin\left(2X_{ij}+\lambda_{f'f;ij}\right)\right)
\end{eqnarray}
for antineutrinos.\\

Eqn.(4) and Eqn.(5) are the oscillation probabilities for $f\neq f'$. We can, also, investigate the
dependence of the survival probability, which is given by $\nu_f\rightarrow\nu_f(\bar\nu_f\rightarrow\bar\nu_f)$,
on parameters of leptonic unitarity quadrangle by taking $f=f'$ and phase shift $\lambda_{f'f;ij}=0$ and calculate disappearance probability. The survival probability $(f=f')$ depends
on four parameters, viz.
\begin{equation}
\left( a_{s}, b_{s}, c_{s}, d_{s}\right)\equiv 
\left(|V_{f0}|^{2},|V_{f1}|^{2},|V_{f2}|^{2},|V_{f3}|^{2}\right),
\end{equation}
which obey the unitarity constraint of the mixing matrix $V$. Hence, under $f=f'$,
Eqn.(4) and Eqn.(5) contains only three independent degrees 
of freedom among all nine parameters in the mixing matrix $V$.
The disappearance probability can be expressed in terms of three independent
parameters $(a_{s},b_{s},c_{s})$ as

\begin{eqnarray}
\nonumber
 P_{dis}=& &1-P_{\nu_{f}\rightarrow \nu_{f}}=4\sum_{i<j}\left(\Re \left(|V_{fi}|^2|V_{fj}|^{2}\right)\sin^{2}\left(X_{ij}\right)\right)-2\sum_{i<j}\left(J_{ff}^{ij}\sin\left(2X_{ij}\right)\right),\\
 \nonumber
        &=&4\Bigl(a_{s}b_{s}\sin^2\left(X_{01}\right)+b_sc_s\sin^2\left(X_{12}\right)\\
        &+&\left(1-a_s-b_s-c_s\right)\left(c_s\sin^2\left(X_{23}\right)+a_s\sin^2\left(X_{03}\right)\right)\Bigr).
  \end{eqnarray}

From Eqn.(7), it is clear that disappearance oscillation experiments cannot provide information
on all the geometric parameters of LUQ. So, in order to investigate $CP$ violation, in a rephase
invariant way, we have to consider neutrino flavor oscillations with non-trivial phase shifts i.e,
$\lambda_{f'f;ij}\neq 0$. For this reason we have considered appearance oscillation probabilities
to possibly determine LUQ parameters.

The orthogonality relation
 \begin{equation}
 V_{f0}V_{f'0}^{*}+V_{f1}V_{f'1}^{*}+V_{f2}V_{f'2}^{*}+V_{f3}V_{f'3}^{*}=0,
 \end{equation}
 
 \begin{figure}[t]
 \begin{center}
 \epsfig{file=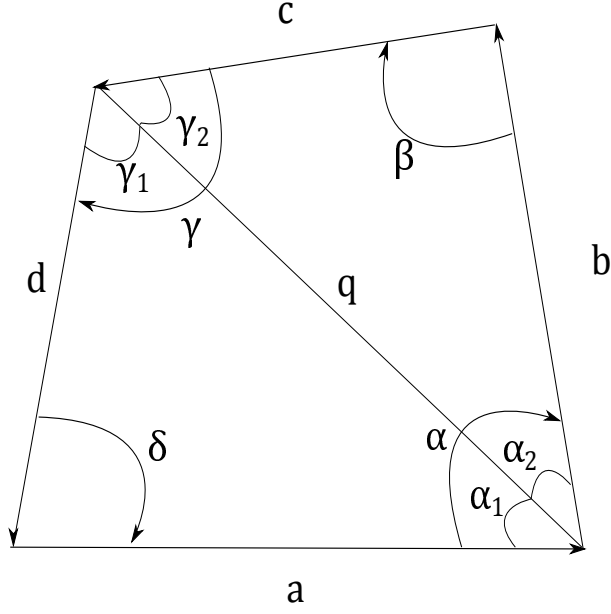,height=8.0cm,width=8.0cm}
 \caption{The leptonic unitarity quadrangle in complex plane. $a, b, c, d$ represent
 the lengths of four sides, and are equal to $\left(|V_{f0}V_{f'0}^{*}|,|V_{f1}V_{f'1}^{*}|,|V_{f2}V_{f'2}^{*}|,|V_{f3}V_{f'3}^{*}|\right)$,
 respectively. $\alpha,\beta,\gamma,\delta$ represent the four angles of the quadrangle.}
 \end{center}
\end{figure}

can be viewed as a quadrangle in complex plane, shown in Fig.(1). Its four sides are,
\begin{equation}
 \left(a,b,c,d\right)=\left(|V_{f0}V_{f'0}^{*}|,|V_{f1}V_{f'1}^{*}|,|V_{f2}V_{f'2}^{*}|,|V_{f3}V_{f'3}^{*}|\right),
\end{equation}
and angles can be expressed as 
\begin{eqnarray}
\nonumber
 \alpha=arg\left(-\frac{V_{f0}V_{f'0}^{*}}{V_{f1}V_{f'1}^{*}}\right),
 \beta=arg\left(-\frac{V_{f1}V_{f'1}^{*}}{V_{f2}V_{f'2}^{*}}\right),\\
 \gamma=arg\left(-\frac{V_{f2}V_{f'2}^{*}}{V_{f3}V_{f'3}^{*}}\right),
 \delta=arg\left(-\frac{V_{f3}V_{f'3}^{*}}{V_{f0}V_{f'0}^{*}}\right).
\end{eqnarray}
From these relations we can write
\begin{equation}
 \alpha=\pi-\lambda_{f'f;01},\beta=\pi-\lambda_{f'f;12},\gamma=\pi-\lambda_{f'f;23},\delta=\pi-\lambda_{f'f;03}.
\end{equation}

Using Eqn.(9) and Eqn.(11), we can write the oscillation probabilities (Eqns.(4)-(5)) in terms 
geometrical parameters of LUQ for neutrino and antineutrino as,
\begin{eqnarray}
\nonumber
P=&&a^{2}+b^{2}+c^{2}+d^{2}-2ab\cos\left(2X_{01}\pm\alpha\right)-2bc\cos\left(2X_{12}\pm\beta\right)\\
  &&-2cd\cos\left(2X_{23}\pm\gamma\right)-2ad\cos\left(2X_{03}\pm\delta\right),
 \end{eqnarray}
where upper (lower) sign in ``$\pm$'' sign correspond to neutrino (antineutrino) oscillations
and $P=P_{\nu_f \rightarrow \nu_{f'}}$ for neutrino $P=P_{\bar\nu_f \rightarrow \bar\nu_{f'}}$ for antineutrino.
Taking into account the fact that $P_{\nu_{f}\rightarrow \nu_{f'}}(L=0)=0$, because neutrinos from the source
do not have sufficient time to oscillate, we can write Eqn.(12) as

\begin{eqnarray}
\nonumber
P=&&4ab \sin\left(X_{01}\pm\alpha\right)\sin X_{01}+4bc\sin\left(X_{12}\pm\beta\right)\sin X_{12}\\
  &&4cd\sin\left(X_{23}\pm\gamma\right)\sin X_{23}+4ad\sin\left(X_{03}\pm\delta\right)\sin X_{03}.
\end{eqnarray}
Eqns.(12)-(13) shows the connection of geometrical parameters $\left(a, b, c, d, \alpha,\beta,\gamma,\delta\right)$
of LUQ to the oscillation probabilities of four neutrino mixing. Here, the geometrical parameters $(\alpha,\beta,\gamma,\delta)$ are playing the role of phase shift in neutrino
flavor oscillations.

\section{$CP$ asymmetry in terms of independent parameters of LUQ}
In order to, uniquely, determine LUQ we choose two sides and three angles viz. $\left(b,c,\alpha,\beta,\gamma\right)$ as the five independent geometrical parameters.
Then, all other parameters of the LUQ can be expressed in 
terms of these five independent parameters. From Fig.(1), we find that
\begin{equation}
\alpha+\beta+\gamma+\delta=2\pi,
\delta=2\pi-\sigma,
\end{equation}
and
\begin{equation}
 a=-q\sin\left(\gamma-\gamma_2\right)\csc\sigma,
 d=-q\sin\left(\alpha-\alpha_2\right)\csc\sigma,
\end{equation}
where, $q=\sqrt{b^2+c^2-2bc\cos\beta}, \gamma_2=\sin^{-1}\left(\frac{b\sin\beta}{q}\right)$, 
$\alpha_2=\sin^{-1}\left(\frac{c\sin\beta}{q}\right)$ and $\sigma\equiv \alpha+\beta+\gamma$.
 We follow the parametrization of the mixing matrix $V$ from \cite{para} and find the independent
 parameters of LUQ viz. $b, c, \alpha, \beta, \gamma$ in terms of mixing angles($\theta_{ij}$) and $CP$-violating
phases($\delta_{ij}$). In general, for n-generations, the mixing matrix $V$ can be parametrized in terms 
of $n_{\theta}=\frac{n(n-1)}{2}$ angles and $n_{\delta}=\frac{n(n+1)}{2}$ phases. However, number of physical 
phases which characterize the mixing matrix is smaller than $n_{\delta}$ because mixing matrix enters into the 
charged current together with fields of charged leptons and neutrinos. We ignore Majorana-type $CP$-violating 
phases as they will not manifest themselves in the oscillation probabilities. Number of Dirac phases in the 
mixing matrix is $n_{\delta}^{D}=\frac{(n-1)(n-2)}{2}$. For $n=4$, $n_{\delta}^{D}=3$. Hence, we write

\begin{eqnarray}
V=V_{23}\tilde{V}_{13}V_{03}V_{12}\tilde{V}_{02}\tilde{V}_{01},
\end{eqnarray}
where $\tilde{V}_{13}, \tilde{V}_{02}, \tilde{V}_{01}$($V_{23}, V_{03}, V_{12}$) are complex(real) elements.     
We can write the independent parameters $b, c, \alpha, \beta, \gamma$ in terms of mixing angles and 
$CP$-violating phases as,

\begin{figure}[t]
 \begin{center}
 \epsfig{file=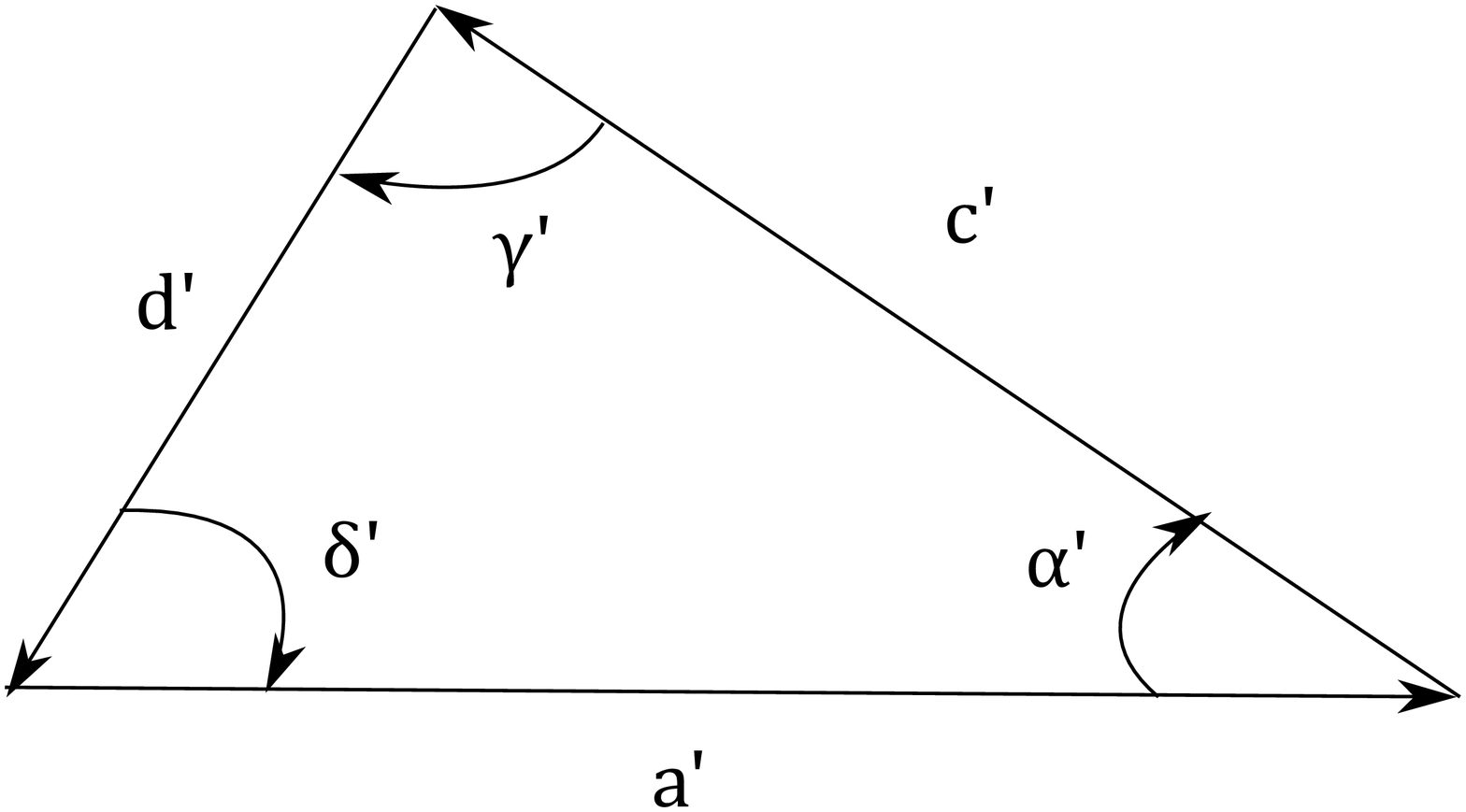,height=5.0cm,width=8.0cm}
\end{center}
 \caption{The leptonic unitarity triangle in complex plane.}
\end{figure}

\small
\begin{eqnarray}
 b=&&|s_{01}^2 \left(c_{13}s_{02}s_{12}+c_{02}s_{03}s_{13} e^{i (\delta_{02}+\delta_{13})}\right)\\
   && \left(c_{02}c_{13}s_{03}s_{23}+e^{i \delta_{02}}s_{02} \left(c_{12}c_{23}-e^{i \delta_{13}}s_{12}s_{13}
   s_{23}\right)\right)|, \nonumber \\
c=&&|\left(-c_{13}s_{02}s_{03}s_{23}+e^{i \delta_{02}} \left(c_{23} (c_{02}c_{12}-s_{12})-e^{i \delta_{13}}s_{13}
   s_{23} (c_{02}s_{12}+c_{12})\right)\right)  \\ &&\left(c_{02}c_{13}s_{12}
   +c_{12}c_{13}-s_{02}s_{03}s_{13} e^{i (\delta_{02}+\delta_{13})}\right)|, \nonumber \\
\alpha=&&\arg\left[-\frac{c_{01}^2}{s_{01}^2}\right],\\ \nonumber
\beta=&&\arg\Bigl[s_{01}^2 \Bigl(c_{13} s_{02} s_{12} + 
     c_{02} e^{i(\delta_{02} + \delta_{13})} s_{03} s_{13}\Bigr) \Bigl(c_{02} c_{13} s_{03} s_{23}+ e^{i \delta_{02}} s_{02}\\
     \nonumber
     &&\left(c_{12} c_{23}-e^{i \delta_{13}} s_{12} s_{13} s_{23}\right)\Bigr)\Bigr]-\arg\Bigl[\Bigl(c_{12} c_{13} + 
     c_{02} c_{13} s_{12} - e^{i (\delta_{02} + \delta_{13})} s_{02} s_{03} s_{13}\Bigr)\\
     &&\Bigl(c_{13} s_{02} s_{03} s_{23} + e^{i \delta_{02}} \left(-c_{02} c_{12} c_{23} + c_{23} s_{12} + 
        e^{i \delta_{13}} \left(c_{12} + c_{02} s_{12}\right) s_{13} s_{23}\right)\Bigr)\Bigr],\\
        \nonumber
  \gamma=&&\arg\Bigl[-e^{-i \left(\delta_{02} + \delta_{13}\right)} \Bigl(-c_{13} \left(c_{12}
        + c_{02} s_{12}\right) + 
    e^{i\left(\delta_{02} + \delta_{13}\right)} s_{02} s_{03} s_{13}\Bigr) \Bigl(c_{13} s_{02} s_{03} s_{23}\\
    \nonumber
   &&+ e^{i \delta_{02}} \left(-c_{02} c_{12} c_{23} + c_{23} s_{12} + 
       e^{i \delta_{13}} (c_{12} + c_{02} s_{12}) s_{13} s_{23}\right)\Bigr)\Bigr]-\arg\Bigl[c_{03}^2 c_{13} s_{13} s_{23}\Bigr], \\   
        \end{eqnarray}

where $s_{ij}=\sin\theta_{ij}$ and $c_{ij}=\cos\theta_{ij}$.
Using Eqns.(14)-(15) we can write, Eqn.(12) in terms of five independent parameters
$(b,c,\alpha,\beta,\gamma)$. The oscillation probability can be written as
\begin{eqnarray}
\nonumber
P=&&q^{2}\sin^{2}\left(\gamma-\gamma_{2}\right)\csc^{2}\sigma+b^{2}+c^{2}+q^{2}\sin^{2}\left(\alpha-\alpha_{2}\right)\csc^{2}\sigma\\
\nonumber
&&+2qb\sin\left(\gamma-\gamma_{2}\right)\csc\sigma\cos\left(2X_{01}\pm\alpha\right)\\
\nonumber
&&-2bc\cos\left(2X_{12}\pm\beta\right)\\
\nonumber
&&+2cq\sin\left(\alpha-\alpha_{2}\right)\csc\sigma\cos\left(2X_{23}\pm\gamma\right)\\
&&-2q^{2}\sin\left(\gamma-\gamma_{2}\right)\sin\left(\alpha-\alpha_{2}\right)\csc^{2}\sigma\cos\left(2X_{03}\mp\sigma\right).
\end{eqnarray} 

For the three generation case, assuming additional mixing angles to be extremely small, the oscillation probability obtained in Eqn.(22), can be written in term of three independent geometric parameters, $c',\alpha' $and $\gamma'$ of unitarity triangle shown in Fig.(2).
Under this approximation and using Eqns.(17)-(21), the expression for oscillation probability(Eqn.(22)) can be written as
\begin{eqnarray}
\nonumber
 P_{3\nu}=&&\frac{c'^{2}}{\sin^{2}\left(\alpha'+\gamma'\right)}\Bigl(\sin^{2}\gamma'+\sin^{2}\left(\alpha'+\gamma'\right)+\sin^{2}\alpha'
 +2\sin\alpha'\sin\left(\alpha'+\gamma'\right)\cos(2X_{23}\pm\gamma')\\
 &&-2\sin\alpha'\sin\gamma'\cos\left(2X_{03}\mp\left(\alpha'+\gamma'\right)\right)\Bigr),
\end{eqnarray}
where, $P_{3\nu}$ is oscillation probability in three generation case with three independent geometric
parameters i.e. one side and two angles, ($c',\alpha'$ and $\gamma')$ of the unitarity triangle, which is same as in Ref.\cite{he}.
Eqn.(22) provide the oscillation probability in terms of five independent geometric parameters of the LUQ. From current neutrino 
oscillation data \cite{garcia}, $|\Delta m_{23}^2|=2.45\times10^{-3} eV^2$, $\Delta m_{12}^2=7.50\times10^{-5} eV^2$ and considering
the case of $E/L\thicksim\Delta m_{12}^2$ we find that $X_{12}$ is $\mathcal{O}(1)$ and $X_{01}, X_{03}>>1$. Thus, the oscillations 
induced by the oscillation frequencies $X_{23}, X_{01}$ and $X_{03}$ will be averaged out due to integration over the neutrino 
production region and energy resolution function. However, the possibility of exploring possible $CP$ asymmetry is still open 
because the combination of results from different oscillation experiments can be used to indirectly measure $CP$ violation effects 
in a $1+3$ scenario. Thus, we can write oscillation probability as
\begin{eqnarray}
\nonumber
 P=&& q^{2}\sin^{2}\left(\gamma-\gamma_{2}\right)\csc^{2}\sigma+b^{2}+c^{2}+q^{2}\sin^{2}\left(\alpha-\alpha_{2}\right)\csc^{2}\sigma\\
 &&-2bc\cos\left(2X_{12}\pm\beta\right).
 \end{eqnarray}
The $CP$ asymmetry can be written as

\begin{eqnarray}
\Delta P=&&4bc\sin\left(2X_{12}\right)\sin\beta.
 \end{eqnarray}
 A realistic method to infer about the parameters $b, c, \alpha, \beta, \gamma$ of the LUQ is to measure the distortion of neutrino energy spectrum. The current neutrino oscillation experiments such as MINOS (with $E/L\thicksim8\times10^{-4} eV^2$) \cite{minos} and NO$\nu$A (with $E/L\thicksim5\times10^{-4}eV^2$) \cite{nova} are insensitive to the oscillation frequency $X_{12}$. However, the condition $E/L\thicksim\Delta m_{12}^2$ can be realized in future neutrino factories \cite{nufact} with $L=(2000-7500)$ km and $E=(1-10) $ GeV. So, one can get the information, only, on three ($b, c, \beta$) out of five independent geometrical parameters ($b, c, \alpha, \beta, \gamma$) of LUQ making it impossible to construct LUQ and to measure of CP asymmetry. 

In general, matter effects becomes important in long baseline experiments. The oscillation probability(Eqn. (22)) will get modified due to interaction of neutrinos with matter. Especially, in Eqn. (22), we note that the independent parameters of LUQ are associated with different oscillating terms. These frequencies($X_{01}, X_{12}, X_{23}, X_{03}$) will have different behavior for a sufficiently large baseline and high precision long baseline experiments, in which case, the oscillation probability(Eqn. 24) will depend on all the five independent geometric parameters of LUQ. Thus, it is possible, in principle, to extract information on all five independent geometric parameters of LUQ in presence of terrestrial matter effects. However, in current LBL experiments such as MINOS and NO$\nu$A, matter effects are insignificant\cite{he} and expression derived for oscillation probability(Eqn. (24)) is still valid. Lets now take the case of short baseline neutrino oscillation experiments, 
with neutrino energy $E=\mathcal{O}(1 GeV)$ and $L=\mathcal{O}(1 km)$ ($E/L\thicksim 0.1 eV^2$ 
such that $|X_{01}|\thicksim |X_{02}|\thicksim |X_{03}|\thicksim 1$).  If the interpretation of the
short baseline anomalies based on neutrino oscillation is correct then we can neglect the oscillations 
due to frequencies $X_{12}, X_{23}$ as their contributions will be small. Under these approximations,
Eqn.(22) can be written as

\begin{eqnarray}
\nonumber
P=&&q^{2}\sin^{2}\left(\gamma-\gamma_{2}\right)\csc^{2}\sigma+b^{2}+c^{2}+q^{2}\sin^{2}\left(\alpha-\alpha_{2}\right)\csc^{2}\sigma\\
\nonumber
&&+2qb\sin\left(\gamma-\gamma_{2}\right)\csc\sigma\cos\left(2X_{01}\pm\alpha\right)\\
&&-2q^{2}\sin\left(\gamma-\gamma_{2}\right)\sin\left(\alpha-\alpha_{2}\right)\csc^{2}\sigma\cos\left(2X_{03}\mp\sigma\right),
\end{eqnarray}

 and $CP$ asymmetry as 
\begin{eqnarray}
\nonumber
 \Delta P=&&-4qb\sin\left(\gamma-\gamma_{2}\right)\csc\sigma\sin\left(2X_{01}\right)\sin\alpha\\
          &&-4q^{2}\sin\left(\gamma-\gamma_{2}\right)\sin\left(\alpha-\alpha_{2}\right)\csc\sigma\sin\left(2X_{03}\right).
\end{eqnarray}
The $CP$ asymmetry is sensitive to all the five independent parameters viz. $b, c, \alpha, \beta, \gamma$ of LUQ. Thus, next generation short baseline experiments provide unique opportunity to construct leptonic unitarity quadrangle and to measure $CP$ violation. Such opportunity to measure $CP$ asymmetry will not be there if we try to directly measure $CP$ phases under aforementioned approximations because the information on $CP$ phases will be lost as new oscillations are, averaged out or small.

\section{Conclusions}
In summary, we have investigated the connection between neutrino survival probability and rephasing invariants of the $4\times4$ neutrino mixing matrix. The $CP$ asymmetry can be measured in two ways. First way is to directly measure $CP$ violating phases in the oscillation experiments and second is to extract information about the parameters of LUQ from oscillation experiments and to construct it. There exist five independent parameters($b, c, \alpha, \beta, \gamma$) of LUQ . We obtain the relation between the oscillation probability, $CP$ asymmetry and independent parameters of LUQ. We have, also, studied the prospects of measuring these parameters and possible measurement of $CP$ asymmetry in future long baseline, neutrino factories and short baseline experiments. We find that it is not possible to fully construct LUQ using data from long baseline experiments because $CP$ asymmetry is sensitive to only three of the five independent parameters of LUQ. However, we expect that matter effects becomes important in long baseline experiments and oscillation probability(Eqn. (22)) will get modified due to interaction of neutrinos with matter. The frequencies($X_{01}, X_{12}, X_{23}, X_{03}$) will have different behavior for a sufficiently large baseline and high precision long baseline experiments, in which case, the oscillation probability(Eqn. 24) will depend on all the five independent geometric parameters of LUQ. Thus, it is possible, in principle, to extract information on all five independent geometric parameters of LUQ in presence of matter effects. Also, using data from short baseline experiments, we can construct and subsequently measure $CP$ violation because $CP$ asymmetry depends on all the five independent geometric parameters of LUQ.

\vspace{1cm}
\textbf{\Large{Acknowledgements}}\\
S. V. acknowledges the  financial support provided by University Grants Commission (UGC)-Basic Science
Research(BSR), Government of India vide Grant No. F.20-2(03)/2013(BSR). S. B. acknowledges the financial 
support provided by the Central University of Himachal Pradesh.

\end{document}